\documentclass[12pt,a4paper]{article}
\usepackage{amsmath}
\usepackage{latexsym}
\usepackage{amssymb}
\usepackage{graphicx,color}
\makeatletter
\def\rddots{\mathinner{\mkern1mu\raise\p@%
    \vbox{\kern7\p@\hbox{.}}\mkern2mu%
    \raise4\p@\hbox{.}\mkern2mu\raise7\p@\hbox{.}\mkern1mu}}
\makeatother
\makeatletter
\def\eqnarray{%
\stepcounter{equation}%
\let\@currentlabel=\theequation
\global\@eqnswtrue
\global\@eqcnt\z@
\tabskip\@centering
\let\\=\@eqncr
$$\halign to \displaywidth\bgroup\@eqnsel\hskip\@centering
$\displaystyle\tabskip\z@{##}$&\global\@eqcnt\@ne
\hfil$\displaystyle{{}##{}}$\hfil
&\global\@eqcnt\tw@$\displaystyle\tabskip\z@{##}$\hfil
\tabskip\@centering&\llap{##}\tabskip\z@\cr}
\makeatother
\begin{document}

\title{\sl A Higher Order Non--Linear Differential Equation \\
and a Generalization of the Airy Function}
\author{
  Kazuyuki FUJII
  \thanks{E-mail address : fujii@yokohama-cu.ac.jp}\\
  Department of Mathematical Sciences\\
  Yokohama City University\\
  Yokohama, 236--0027\\
  Japan
  }
\date{}
\maketitle
%
%
%
%
\begin{abstract}
  In this paper a higher order non--linear differential equation is 
  given and it becomes a higher order Airy equation (in our terminology) 
  under the Cole--Hopf transformation. 
  
  For the even case a solution is explicitly constructed, which is 
  a generalization of the Airy function.
\end{abstract}
%


%
%
%
%

We start with an example to state our motivation. For the simple 
non--linear equation of Riccati type
\begin{equation}
\label{eq:non-lenear equation}
y{'}+y^{2}=x
\end{equation}
with a smooth function $y=y(x)$, we apply the Cole--Hopf transformation 
\begin{equation}
\label{eq:Cole-Hopf}
y=\frac{d}{dx}\log{u}=\frac{u{'}}{u}\quad \left(u=u(x)\right)
\end{equation}
to (\ref{eq:non-lenear equation}). Then we obtain the famous Airy 
equation \cite{Airy}
\begin{equation}
\label{eq:Airy equation}
u{''}=xu.
\end{equation}

This equation plays an important role in both Quantum Optics and 
Mathematical Physics, so many studies have been made. 
See for example \cite{Schulman}. 

This has two well--known solutions called the Airy function $Ai(x)$ 
and the Airy function of the second kind $Bi(x)$. 
For the details see \cite{Weisstein}. 
In particular, $Ai(x)$ is written in terms of an improper integral
\begin{equation}
\label{eq:Airy function}
u(x)=\frac{1}{\pi}
\int_{0}^{\infty}\cos\left(\frac{t^{3}}{3}+xt\right)dt.
\end{equation}
Therefore, a solution of (\ref{eq:non-lenear equation}) is given by
\[
y=\frac{-\int_{0}^{\infty}{t}\sin\left(\frac{t^{3}}{3}+xt\right)dt}
       {\int_{0}^{\infty}\cos\left(\frac{t^{3}}{3}+xt\right)dt}.
\]
The formula (\ref{eq:Airy function}) is very suggestive and will be 
generalized in the following.

Next, we would like to generalize the Airy equation 
(\ref{eq:Airy equation}). One of candidates is
\begin{equation}
\label{eq:generalized Airy equation}
u^{(n)}\equiv \frac{d^{n}}{dx^{n}}u=xu \quad\mbox{for}\quad  n\geq 2
\end{equation}
and we adopt this in the paper. 

In the following we call this equation a higher order Airy equation. 
Then our question is as follows : What is a non--linear equation like 
(\ref{eq:non-lenear equation}) corresponding to the higher order 
Airy equation (\ref{eq:generalized Airy equation}) ? 

After some consideration we reach the following equation. For a 
given smooth function $y$ we set a sequence of functions 
$\{f_{n}\}_{n\geq 1}$ with
\[
f_{n}\equiv f_{n}(y,y{'},y{''},\cdots,y^{(n-1)})
\]
recurrently like
\begin{equation}
\label{eq:generalized non-linear function}
f_{n+1}=\left(\frac{d}{dx}+y\right)f_{n}\quad\mbox{and}\quad f_{1}=y.
\end{equation}
This $f_{n}$ is also written in a compact form
\begin{equation}
\label{eq:generalized non-linear function 2}
f_{n}=\left(\frac{d}{dx}+y\right)^{n-1}y.
\end{equation}

A few examples are
\begin{eqnarray*}
f_{2}&=&y{'}+y^{2},\\
f_{3}&=&y{''}+3yy{'}+y^{3},\\
f_{4}&=&y^{3}+4yy{''}+3y{'}^{2}+6y^{2}y{'}+y^{4}.
\end{eqnarray*}

Therefore our higher order non--linear differential equation is 
defined by
\begin{equation}
\label{eq:generalized non-linear equation}
f_{n}(y,y{'},y{''},\cdots,y^{(n-1)})=x
\end{equation}
for $n\geq 2$. For example,
\begin{eqnarray*}
&&y{'}+y^{2}=x,\\
&&y{''}+3yy{'}+y^{3}=x,\\
&&y^{3}+4yy{''}+3y{'}^{2}+6y^{2}y{'}+y^{4}=x.
\end{eqnarray*}

Here we state one of the results. 

\par \noindent
{\bf Result I}\ \ Under the Cole--Hopf transformation 
(\ref{eq:Cole-Hopf}) ($y=u{'}/u$) we have
\begin{equation}
\label{eq:generalized Cole-Hopf}
f_{n}(y,y{'},y{''},\cdots,y^{(n-1)})=\frac{u^{(n)}}{u}.
\end{equation}

The proof is easy and due to the mathematical induction. We assume 
\[
f_{n}(y,y{'},y{''},\cdots,y^{(n-1)})=\frac{u^{(n)}}{u}
\]
and operate $d/dx +y$ to the both sides. Then
\[
f_{n+1}=\left(\frac{d}{dx}+y\right)f_{n}=
\left(\frac{d}{dx}+\frac{u{'}}{u}\right)\frac{u^{(n)}}{u}=
\frac{u^{(n+1)}}{u}.
\]

\vspace{3mm}
A comment is in order.\ \ We would like to call the transformation
\[
y=\frac{u^{(n)}}{u}\quad \mbox{for}\quad n\geq 2
\]
a higher order Cole--Hopf transformation. 

Therefore our equation (\ref{eq:generalized non-linear equation}) 
becomes the higher order Airy equation
\[
u^{(n)}=xu
\]
in (\ref{eq:generalized Airy equation}).

Next, let us construct some solution like (\ref{eq:Airy function}) 
to the equation (\ref{eq:generalized Airy equation}).

\par \noindent
{\bf Result II}\ \ We set $m\geq 1$.

\par \noindent
(a)\ \ For $n=4m-2$\ a solution is given by
\begin{equation}
\label{eq:higher order Airy function 1}
u(x)=\frac{1}{\pi}
\int_{0}^{\infty}\cos\left(\frac{t^{4m-1}}{4m-1}+xt\right)dt.
\end{equation}
\par \noindent
(b)\ \ For $n=4m$\ a solution is given by
\begin{equation}
\label{eq:higher order Airy function 2}
u(x)=\frac{1}{\pi}
\int_{0}^{\infty}\cos\left(\frac{t^{4m+1}}{4m+1}-xt\right)dt.
\end{equation}

The proof is standard. For example, 
(\ref{eq:higher order Airy function 2}) is proved as follows. 
\begin{eqnarray*}
u^{(4m)}
&=&\frac{1}{\pi}
\int_{0}^{\infty}t^{4m}\cos\left(\frac{t^{4m+1}}{4m+1}-xt\right)dt
\\
&=&\frac{1}{\pi}
\int_{0}^{\infty}(t^{4m}-x+x)
\cos\left(\frac{t^{4m+1}}{4m+1}-xt\right)dt \\
&=&\frac{1}{\pi}
\int_{0}^{\infty}(t^{4m}-x)\cos\left(\frac{t^{4m+1}}{4m+1}-xt\right)dt
+xu \\
&=&\frac{1}{\pi}
\int_{0}^{\infty}\frac{d}{dt}\sin\left(\frac{t^{4m+1}}{4m+1}-xt\right)dt
+xu \\
&=&xu.
\end{eqnarray*}

A comment is in order.\ \ For $n=2m+1$ we could not construct a solution 
like (\ref{eq:higher order Airy function 1}) or 
(\ref{eq:higher order Airy function 2}), so the construction is left to 
readers.

Finally, we compute initial values of the solution 
(\ref{eq:higher order Airy function 1}) and 
(\ref{eq:higher order Airy function 2}). For simplicity we set
\begin{equation}
\label{eq:}
v_{\pm}(x)=\frac{1}{\pi}
\int_{0}^{\infty}\cos\left(\frac{t^{n+1}}{n+1}\pm xt\right)dt
\end{equation}
and compute $v_{\pm}^{(k)}(0)$ for $0\leq k \leq n-1$. Since
\[
v_{\pm}^{(k)}(x)=(\pm)^{k}\frac{1}{\pi}
\int_{0}^{\infty}t^{k}\cos\left(\frac{t^{n+1}}{n+1}\pm xt+
\frac{k\pi}{2}\right)dt
\]
and
\[
v_{\pm}^{(k)}(0)=(\pm)^{k}\frac{1}{\pi}
\int_{0}^{\infty}t^{k}\cos\left(\frac{t^{n+1}}{n+1}+\frac{k\pi}{2}
\right)dt
\]
the change of variables $s=t^{n+1}/(n+1)\Leftrightarrow 
t=(n+1)^{\frac{1}{n+1}}s^{\frac{1}{n+1}}$ gives
\[
\int_{0}^{\infty}t^{k}\cos\left(\frac{t^{n+1}}{n+1}+\frac{k\pi}{2}
\right)dt
=
(n+1)^{\frac{k+1}{n+1}-1}\int_{0}^{\infty}
s^{\frac{k+1}{n+1}-1}\cos\left(s+\frac{k\pi}{2}\right)ds.
\]
By the Mellin transformation we have
\[
\int_{0}^{\infty}t^{k}\cos\left(\frac{t^{n+1}}{n+1}+\frac{k\pi}{2}
\right)dt
=
(n+1)^{\frac{k+1}{n+1}-1}\Gamma\left(\frac{k+1}{n+1}\right)
\cos\left(\frac{k+1}{2(n+1)}\pi+\frac{k\pi}{2}\right).
\]
As a result
\begin{equation}
v_{\pm}^{(k)}(0)=(\pm)^{k}\frac{1}{\pi}
(n+1)^{\frac{k+1}{n+1}-1}\Gamma\left(\frac{k+1}{n+1}\right)
\cos\left(\frac{k+1}{2(n+1)}\pi+\frac{k\pi}{2}\right).
\end{equation}

Moreover, let us change to the usual form. By the well--known 
formula
\[
\Gamma(p)\Gamma(1-p)=\frac{\pi}{\sin(p\pi)}\quad (0<p<1)
\]
(see for example \cite{WW}) we finally obtain

\par \noindent
{\bf Result III}
\begin{equation}
v_{\pm}^{(k)}(0)
=
\frac{(\pm)^{k}}{(n+1)^{\frac{n-k}{n+1}}\Gamma\left(\frac{n-k}{n+1}\right)}
\frac{\cos\left(\frac{k+1}{2(n+1)}\pi+\frac{k\pi}{2}\right)}
     {\sin\left(\frac{k+1}{n+1}\pi\right)}
\end{equation}
for $0\leq k\leq n-1$.

For example, for $n=2$ and $k=0,\ 1$
\[
v_{+}^{}(0)=\frac{1}{3^{\frac{2}{3}}\Gamma(\frac{2}{3})},\quad 
v_{+}{'}(0)=-\frac{1}{3^{\frac{1}{3}}\Gamma(\frac{1}{3})}.
\]

\vspace{3mm}
Since initial values of the solution has been determined we determine 
the asymptotic behaviours.

\par \noindent
{\bf Result IV}\ \ We set $m\geq 1$.\ \ 
\par \noindent (I) For $x\gg 0$
\begin{equation}
\label{eq:asymptotic behaviours 1}
u(x)\sim
\frac{e^{-\frac{2m}{2m+1}x^{\frac{2m+1}{2m}}}}{\sqrt{\pi}\sqrt{4m}x^{\frac{2m-1}{4m}}},
\end{equation}

\par \noindent (II) For $x\ll 0$
\begin{equation}
\label{eq:asymptotic behaviours 2}
u(x)\sim
\frac{1}{\sqrt{\pi}\sqrt{m}\left(-x \right)^{\frac{2m-1}{4m}}}
\sum_{k=0}^{m-1}e^{\alpha\cos\left(\frac{1+2k}{2m}\pi\right)}
\sin\left(\alpha\sin\left(\frac{1+2k}{2m}\pi\right)+\frac{1+2k}{4m}\pi\right)
\end{equation}
where $\alpha=\frac{2m}{2m+1}\left(-x \right)^{\frac{2m+1}{2m}}$.

The proof is left to readers.

\vspace{5mm}
In this paper we gave the higher order non--linear differential 
equation, which is a generalization of the equation of Riccati 
type and showed that it became the higher order Airy equation 
under the Cole--Hopf transformation. 

We constructed some solution for the even case and also calculated 
initial values and asymptotic behaviours of the solution.

Our work may be more or less known. However, the author could not 
find such a reference.

The work is only in the first stage and there are many tasks 
left to readers. See and study \cite{Weisstein} in more detail 
from the view point of the paper.

We believe that our generalization is natural. However, it is not 
clear at the present time whether it is useful or not in Mathematical 
Physics or Quantum Optics. This is our future task.


\end{document}